\documentclass[default,iicol]{sn-jnl}

\jyear{2021}%

\theoremstyle{thmstyleone}%
\theoremstyle{thmstyletwo}%

\theoremstyle{thmstylethree}%

\usepackage{bm}
\usepackage{accents}
\usepackage{ulem}

\usepackage{amsmath}
\usepackage{bm}

\newcommand{\rr}{\bm{r}}

\newcommand{\Ra}{\rm{Ra}}
\newcommand{\Nu}{\rm{Nu}}

\newcommand{\uu}{\bm{u}}

\newcommand{\nab }{\mathbf{\nabla}}

\jyear{2022}%
\begin{document}

\title[Reconstruction of Rayleigh-B\'ernard Flows]{Reconstructing
Rayleigh-B\'enard flows out of temperature-only measurements using Physics-Informed Neural Networks}

\author*[1]{\fnm{Patricio} \sur{Clark Di Leoni}}\email{pclarkdileoni@udesa.edu.ar}
\author[2,3]{\fnm{Lokahith} \sur{Agasthya}}\email{lnagasthya@gmail.com}
\author[2]{\fnm{Michele} \sur{Buzzicotti}}\email{michele.buzzicotti@roma2.infn.it}
\author[2]{\fnm{Luca} \sur{Biferale}}\email{luca.biferale@roma2.infn.it}

\affil[1]{\orgdiv{Departmento de Ingeniería}, \orgname{Universidad de San Andrés},
\orgaddress{\city{Buenos Aires}, 
\country{Argentina}}}

\affil[2]{\orgdiv{Department of Physics and INFN}, \orgname{University of Rome ``Tor Vergata''},
\orgaddress{\city{Rome}, 
\country{Italy}}}

\affil[3]{Institute of Science and Technology Austria,
\orgaddress{\street{Am Campus 1}, \city{Klosterneuburg}, \postcode{3400}, \country{Austria}}}
 


\abstract{
We investigate the capabilities of Physics-Informed Neural Networks (PINNs) to reconstruct turbulent Rayleigh-B\'enard flows using only temperature information. We perform a quantitative analysis of the quality of the reconstructions at various amounts of low-passed-filtered information and turbulent intensities. We compare our results with those obtained via nudging, a classical equation-informed data assimilation technique. At low Rayleigh numbers, PINNs are able to reconstruct with high precision, comparable to the one achieved  with nudging. At high Rayleigh numbers, PINNs outperform nudging and are able to achieve satisfactory reconstruction of the velocity fields only when data for temperature is provided with high spatial and temporal density. When data becomes sparse, the PINNs performance worsens, not only in a point-to-point error sense but also, and contrary to nudging, in a statistical sense, as can be seen in the probability density functions and energy spectra.}


\maketitle

\section{Introduction}
\label{sec:introduction}
Understanding the type and quantity of information needed to reconstruct the state of a physical system carries important implications for both its fundamental study and its real-world applications. In this work we analyze this question for the case of thermally driven flows. Thermally driven flows are at the core of several geophysical and industrial systems such as, atmospheric convection, \cite{hartmann2001tropical,suselj2019unified}, oceanic convection \cite{vaage2018ocean}, mantle convection \cite{kronbichler2012high} and pure-metal melting \cite{brent1988enthalpy}. These flows can exhibit a wide variety of behaviors and structures, ranging from plume formation to fully developed turbulence \cite{Grossman-Lohse-Scaling}. It was first conjectured by Charney \cite{ch03600a} that temperature measurements alone are enough to reconstruct the whole state of the atmosphere. This conjecture has been studied both theoretically and numerically in simple convective systems \cite{ghil_balanced_1977,ghil_time-continuous_1979}, 3D Planetary Geostrophic models \cite{farhat2016charney}, and Rayleigh-B\'enard flows in non-turbulent regimes \cite{altaf_downscaling_2017, hammoud_cdanet_2022}, in the infinite Prandtl number limit \cite{farhat2020data}, and at moderate and high Rayleigh numbers \cite{agasthya_reconstructing_2022}. These studies showed the importance of setting a correct velocity prior  to get a good reconstruction \cite{altaf_downscaling_2017}, and the fragility to get a time-independent full synchronization at high Rayleigh numbers  \cite{hammoud_cdanet_2022,agasthya_reconstructing_2022}. A deeper understanding of the Charney conjecture is then important not only to elucidate the interplay between velocity and temperature but also to improve current forecasts and Data Assimilation \cite{Kalnay,Bauer15} schemes.
The aim of this paper is to continue the work presented in \cite{agasthya_reconstructing_2022} in 2D turbulent Rayleigh-B\'enard flows. Whilst the original work used Nudging \cite{Lakshmivarahan13,buzzicotti2020synchronizing,clark_di_leoni_synchronization_2020} , a synchronization equations-based DA tool, to reconstruct the flows, we now use Physics-Informed Neural Networks (PINNs). PINNs are neural networks designed to approximate the solution of systems of partial differential equations \cite{raissi_physics-informed_2019}. They have been used in inverse problems with partial information  \cite{raissi_hidden_2020, shukla_physics-informed_2020}, to reconstruct turbulent flows out of measurements \cite{cai_flow_2021, wang_dense_2022, du_VarPINN2022, clark_di_leoni_reconstructing_2022}, and to assimilate statistical data into synthetically generated fields \cite{angriman_generation_2022}. For detailed reviews on PINNs, see \cite{cai_physics-informed_2021, cuomo_scientific_2022}. Using a flow coming from a Direct Numerical Simulation of the 2D Rayleigh-B\'enard system at two different Rayleigh numbers, one moderate and one high, we use PINNs to perform reconstructions using varying amounts of data, characterized by the separation distance between measuring probes. We show that PINNs are successful at the task, even at high Rayleigh numbers, where the correlations between the temperature field (for which we provide information) and the velocity fields (for which we do not) diminish.

The paper is organized as follows, in Sec.~\ref{sec:rb_data_gen} we introduce the Rayleigh-B\'enard equations and describe the data generation procedures while in Sec.~\ref{sec:pinn_exps} we describe the PINN technique and how it is applied in this context. Results are presented in Sec.~\ref{sec:results} and conclusions in Sec.~\ref{sec:conclusions}.
\begin{figure}
\includegraphics[width=0.5\textwidth]{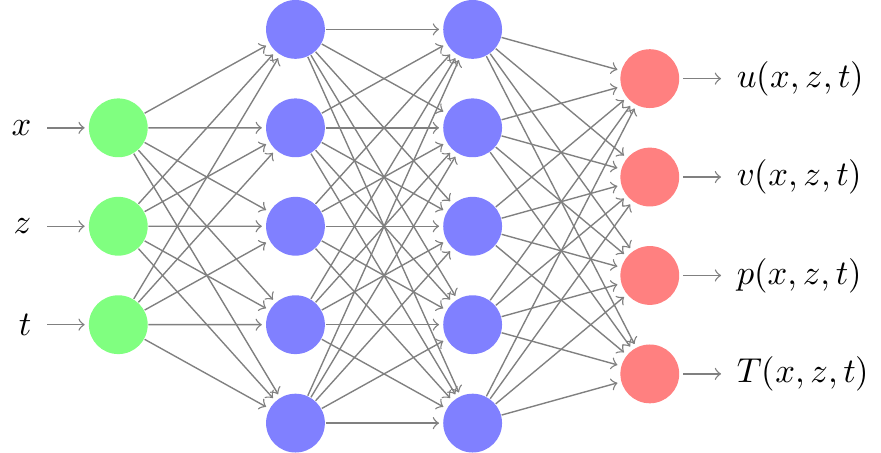}
    \caption{Diagram of the PINN used.}
    \label{fig:pinn_diag}
\end{figure}
\section{Methods}

\subsection{Rayleigh-B\'enard flows and data generation}
\label{sec:rb_data_gen}

Rayleigh-B\'enard convection consists of a planar horizontal layer of fluid that is heated from below with respect to gravity. If density fluctuations are small, the Boussinesq approximation may be employed and the fluid can be described in terms of an incompressible velocity field plus a temperature field. Taking $\uu = (u,v)$ to be the horizontal and vertical components of the velocity field, respectively, $T$ the temperature and $p$ the pressure, in a 2D geometry and with the average temperature set to zero and the density to unity the equations take the form

\begin{gather}\label{eq:R-B-eqns}
    \partial_t \uu + (\uu \cdot \nab) \uu  =  -\nab p  + \nu \nabla^2 \uu -
    \beta T  g \hat{z}, \\   \label{eq:R-B-eqns1}
    \frac{\partial T}{\partial t} + \uu \cdot \nab T = \kappa \nabla^2 T, 
\end{gather}
where $\beta$ is the thermal expansion coefficient of the fluid, $\nu$ is its
kinematic viscosity, $\kappa$ its thermal conductivity and $g\hat{z}$ the
acceleration due to gravity. The domain is $L_x$ wide and $L_z$ tall and has
periodic boundary conditions in the horizontal direction $\hat{x}$. At the top
and bottom boundaries, the boundary conditions are

\begin{equation}\label{eq:R-B_BoundaryCs}
\begin{gathered}
   T(z=0) = T_d,\qquad
   T(z=L_z) = -T_d, \\
   \uu(z=0) = \uu(z=L_z) = 0. \\
\end{gathered}
\end{equation}
where $T_d>0$. The characteristic velocity scale is given by $u_0 = \sqrt{g L_z \beta 2 T_d}$ and the turnover time by $\tau_0 = 2 L_z/u_0$. Several dimensionless numbers can be used to describe Rayleigh-B\'enard flows. The Rayleigh number gives a measure of the ratio between buoyant and viscous forces and is given by

\begin{equation}
    \Ra = g \beta \frac{2 T_d L_z^3}{\nu \kappa}.
\end{equation}
The Prandtl number is the ratio between momentum diffusivity and thermal diffusivity, namely

\begin{equation}
    \Pr = \frac{\nu}{\kappa}.
\end{equation}
The Nusselt number measures the ratio of heat transfer due to convection versus that due to conduction and is given by
\begin{equation}
    \Nu = \frac{\langle v T - \kappa \partial_z T \rangle}{\kappa 2 T_d / L_z},
\end{equation}
where $\langle . \rangle$ indicate the ensemble average over the whole domain. Finally, as an estimate of the size of the smallest scales of the system we define the Kolmogorov length scale
$
\eta_\kappa = (\nu^3/\epsilon)^{1/4},
$
where $\epsilon = (\nu \kappa^2 / L^4_z) (\Nu - 1) \Ra$ is the average rate of energy dissipation \cite{siggia_high_1994}.

The reference, or ground-truth, flow of our numerical experiments was produced by evolving Eqs.~\eqref{eq:R-B-eqns}-\eqref{eq:R-B-eqns1} using a Lattice-Boltzmann method at two different Rayleigh numbers, $\Ra_1 = 7.2 \times 10^7$ and $\Ra_2 = 36.3 \times 10^7$. The reference flows are denoted by variables $u^r$, $v^r$, $p^r$, and $T^r$. Details on the numerical method can be found in \cite{agasthya_reconstructing_2022}.
\begin{figure*}
\includegraphics[width=0.95\textwidth]{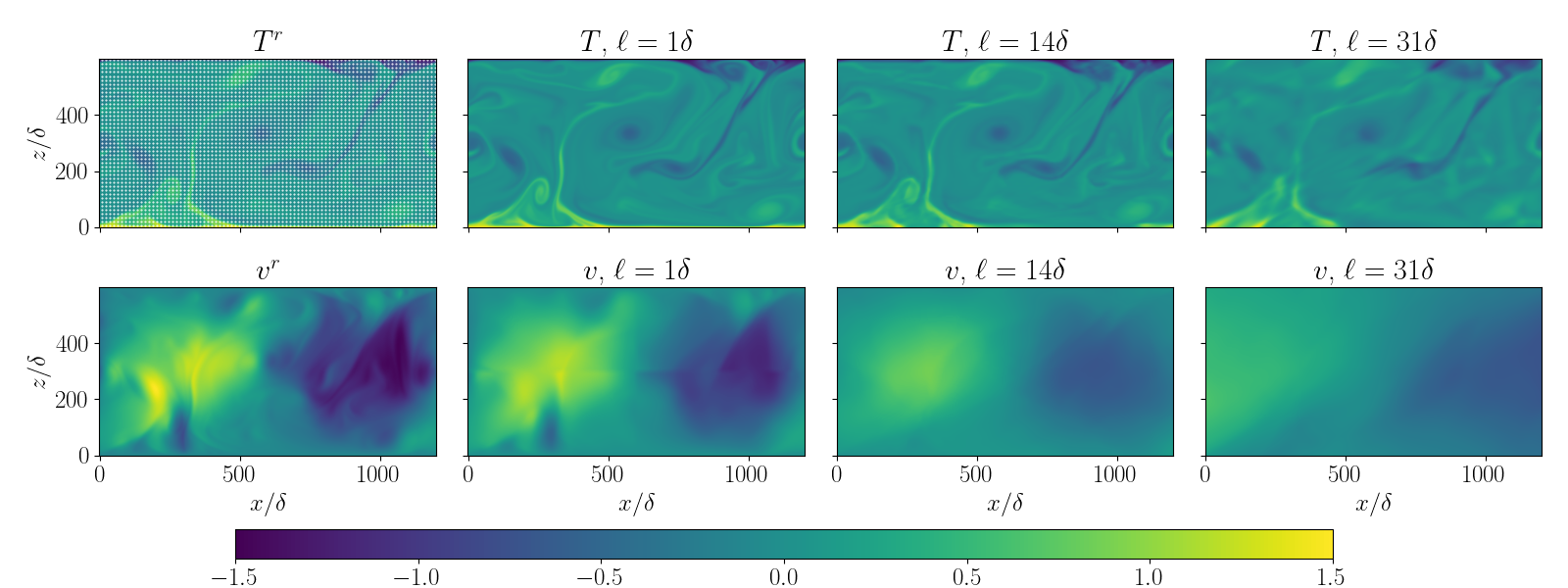}
    \caption{Visualizations of temperature (top) and vertical velocity (bottom) for the flow with $\Ra_2$. The left column shows the reference data, the other three columns show the reconstructions obatined with $\ell/\delta = 1$, $14$ and $31$. The locations of the measuring probes (corresponding to the case with $\ell=14\delta$) are marked with white dots on top of $T^r$. All visualizations share the same colorbar.}
    \label{fig:viz}
\end{figure*}
In the first case $L_x = 864 \delta$, $L_z = 432\delta$, $T_d = 2.5 T_0$, and $u_0 = 1.31 U_0$, while in the second case $L_x = 1200 \delta $, $L_z = 600 \delta$, $T_d = 1.5 T_0$, and $u_0 = 2.12 U_0$. In both cases $\nu = 6.67 \times 10^{-4}$, $\Pr = 1$, $\delta=1$, $T_0 = 1/100$,  $U_0 = 1/100$, and $\eta_\kappa \approx 2 \delta$. In the first case $\Nu \approx 25$, while in the second case $\Nu \approx 39$. The flows were first allowed to reach a statistically stationary state and then data was extracted on a equally spaced rectangular grid of separation distance $\ell \in [1:31]\delta$ at specific time intervals. In the first case, data was extracted over 8 separate time windows 11 snapshots long at a sampling rate of $164/\tau_0$. In the second case, data was extracted over 10 separated time windows also 11 snapshots long but at a sampling rate of $1130/\tau_0$, as this flow contains faster scales than the other one. Data was extracted over the whole spatial domain, except for the case with $\ell=1$ at $\Ra_2$, where the domain was split into four quadrants, thus necessitating four PINNs for the reconstruction of the whole domain. Each dataset $\Omega_d$ is then identified by the Rayleigh number of the flow, the grid spacing $\ell$, and the temporal window. The sets of collocation points $\Omega_p$ where the physics part of the loss function is evaluated (see below) consist of all points where data part are evaluated, i.e. $\Omega_d$, plus randomly selected points not necessarily lying on the $\ell$-spaced sampling grid. Note that as no data are used when evaluating the physics part of the loss function, we are not increasing the density of data used. The number of extra collocation points was 0, 0, 1, 2, 3, and 4 for every spatial position  in $\Omega_d$ for $\ell/\delta=1, 7, 10, 14, 22$, and $31$,  respectively. To evaluate the PINNs' performances testing datasets $\Omega_t$ consisting of the full fields, i.e., not subsampled in space, were used.

\subsection{Flow reconstruction using PINNs}
\label{sec:pinn_exps}
Figure~\ref{fig:pinn_diag} shows a diagram of the PINN used in our experiments. The PINN takes coordinates $(x, z, t)$ as input and outputs fields $(u, v, p, T)$ at the specified coordinate. All PINNs presented here are five layers deep, have 100 hidden units in each layer and use ELU as activation functions. The loss function used to train them has the form
\begin{align*}
    L = L_d + \lambda L_p,
\end{align*}
with the contribution given by the measured data is given by
\begin{align*}
    L_d = \frac{1}{N(\Omega_d) T^2_0} \sum_{j \in \Omega_d} \vert T(x_j, z_j, t_j) - T^{r}_j \vert^2,
\end{align*}
and the one given by the imposition of the correct equations of motion by:
\begin{align*}
    L_p =  \frac{\delta^2}{N(\Omega_p)} \sum_{j \in \Omega_p} & \left( \lambda_u U^{-4}_0 f_u +
    \lambda_T T^{-2}_0 U^{-2}_0 f_T + \right.
    \\
    & \left. \lambda_i U^{-2}_0 f_i \right),
\end{align*}
with
\begin{align*}
    f_u =
    \vert \partial_t \uu + (\uu \cdot \nab) \uu  + \nab p  - \nu \nabla^2 \uu + \beta T  g \hat{z} \vert^2,
\end{align*}
\begin{align*}
    f_T =  \vert \partial_t T + \uu \cdot \nab T - \kappa \nabla^2 T \vert^2,
\end{align*}
and
\begin{align*}
    f_i = \vert \nabla \cdot \uu \vert^2, 
\end{align*}
and where $\lambda$, $\lambda_u$, $\lambda_T$ and $\lambda_i$ are extra hyperparameters set fixed to $10^5$, $1$, $10^{-1}$ and $10^3$, respectively, throughout. The subsets of points $\Omega_d$ and $\Omega_p$ where the summations are evaluated are explained in the section above, $N(\Omega_d)$ and $N(\Omega_p)$ denote the number of point in the respective datasets. The PINNs were trained using the Adam algorithm with a learning rate of $10^{-4}$ for 60000 epochs. 

\subsection{Flow reconstruction using Nudging}
\label{sec:pinn_exps}

Nudging is an equation-informed data assimilation tool, using the evolution of the Navier-Stokes equations \eqref{eq:R-B-eqns} supplemented by a Newton relaxation term proportional to $ \alpha ( T_(x_j,z_j,t_j)^r-T) $ in the temperature evolution. The resulting equations take the form

\begin{gather}\label{eq:nudging-eqns}
    \partial_t \uu + (\uu \cdot \nab) \uu  =  -\nab p  + \nu \nabla^2 \uu -
    \beta T  g \hat{z}, \\   
    \frac{\partial T}{\partial t} + \uu \cdot \nab T = \kappa \nabla^2 T - \alpha \mathcal{I} (T^r - T), \nonumber
\end{gather}
where $\alpha$ is the amplitude of the nudging term and has units of frequency and $\mathcal{I}$ is a filtering operator equal to 1 where the data is available, i.e. the measuring probes, and zero otherwise. The idea in nudging is to force the equation of motion to ``follow'' the available information where it is available and leave the equations of motion to refill the gaps in the whole space-time domain. For further details about the optimal selection for the nudging parameter $\alpha$ and how to interpolate in time and space the supplied data see  \cite{agasthya_reconstructing_2022}. In contrast with PINNs, Nudging naturally enforces all physical constraints everywhere, at the cost  of evolving the partial differential equations on the entire domain.

\subsection{Error assessment}
In order to assess the reconstructions obtained, we define the point-to-point error field for temperature and velocity as:
\begin{eqnarray}
   & T_\Delta(\rr,t) =  T^r(\rr,t) - T(\rr,t);\nonumber \\
   & v_\Delta(\rr,t) =   v^r(\rr,t) - v(\rr,t).
\end{eqnarray}
Global normalized errors are then given by

\begin{equation}
    \Delta_T  = \frac{\langle T_\Delta^2(\rr,t) \rangle}{\langle T^2(\rr,t) \rangle};\qquad 
    \Delta_v = \frac{\langle v_\Delta^2(\rr,t) \rangle}{\langle v^2(\rr,t) \rangle},
\end{equation}

where $\langle \cdot \rangle$ indicates the average over the entire domain. It is important to stress that no information on the temperature at the boundary was provided (similarly to what implemented for nudging). 

The scale-by-scale analysis is performed by analyzing the energy spectra:

\begin{equation}
    E_f (k) = \langle \vert \hat{f} (k, z_0, t) \vert^2 \rangle_t,
\end{equation}
where $f$ is the field studied (either $v$, $v_\Delta$, $T$, or $T_\Delta$), $\hat{f}(k, z_0, t)$ are the Fourier coefficients of $f$ calculated along the horizontal direction at position $z_0 = L_z/2$ and time $t$, and $\langle , \rangle_t$ denotes the time average. 
\section{Results}
\label{sec:results}

\begin{figure}
    \includegraphics[width=0.45\textwidth]{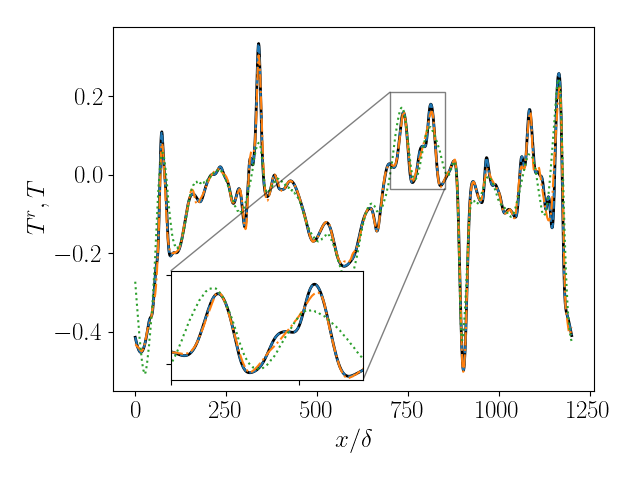}
    \includegraphics[width=0.45\textwidth]{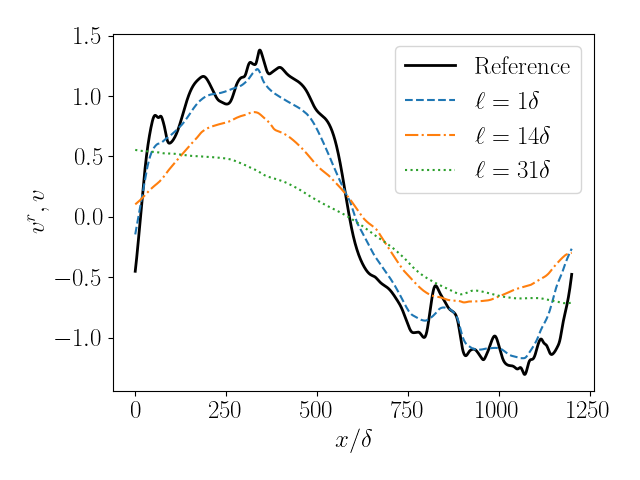}
    \caption{Temperature (top) and velocity (bottom) profiles for the case with $\Ra_2$. The reconstructions shown were obtained using $\ell/\delta=1$, 14, and 31. Both figures share the same legend.}
    \label{fig:prof_ra2}
\end{figure}

Figure~\ref{fig:viz} shows visualizations of the reference temperature $T^r$ and vertical velocity $v^r$ of the flow with $\Ra_2$ at a given time and their PINN-reconstructed counterparts, $T$ and $v$, for three reconstructions, one performed with $\ell/\delta=1$, one with $\ell/\delta=14$ and one with $\ell/\delta=31$. The location of the temperature measuring probes (corresponding to the $\ell/\delta=14$ case) are marked with white dots. These visualizations not only show the characteristics of the flow, but also give a qualitative glimpse into the overall results: PINNs are able to reconstruct/infer  the velocity field using only temperature information, even at high Rayleigh number, but with clear difficulties to capture small-scales fluctuations, as shown by the blurred velocity configurations (middle and right bottom panels). To further stress this point, we show the  horizontal profile at $z_0=300 \delta$ in Fig.~\ref{fig:prof_ra2} with three different  $\ell/\delta=1,14,31$ for temperature (top), and  velocity (bottom). As seen, the temperature reconstruction almost overlaps with the reference data for all cases, except the one with $\ell/\delta=31$. The reconstructed velocity field, on the other hand, is able to correctly match the large scale structure of the reference flow, but is missing smaller scale features. Furthermore,  the effects of not enforcing periodic boundary conditions along the horizontal direction can be clearly seen in the $\ell/\delta=31$ case.

\begin{figure}
    \includegraphics[width=0.45\textwidth]{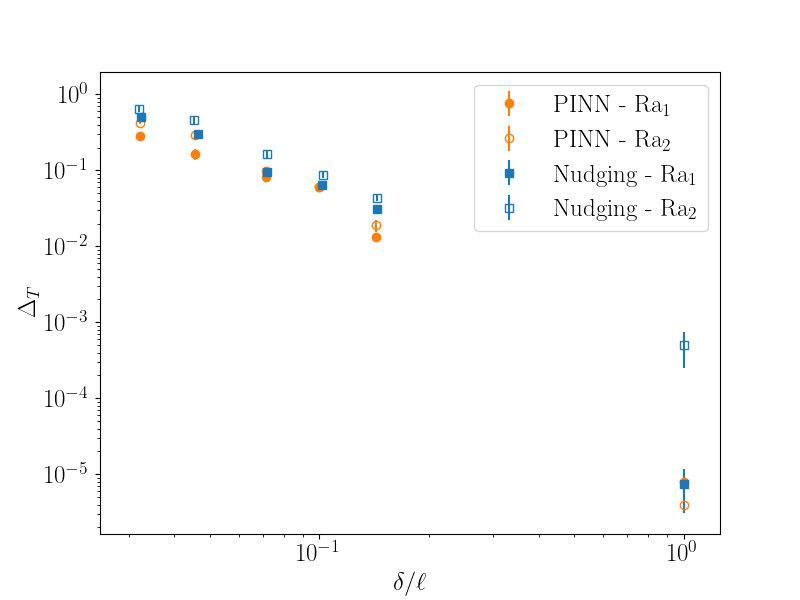}
    \includegraphics[width=0.45\textwidth]{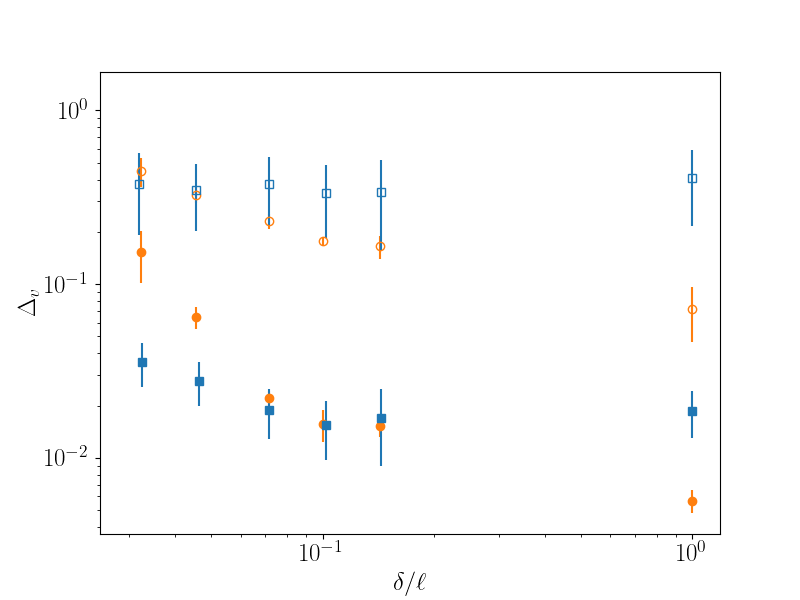}
    \caption{Global temperature errors $\Delta_T$ (top) and global velocity errors $\Delta_v$ (bottom) as a function of $\delta/\ell$. Full markers denote the results for the case with $\Ra_1$, while empty markers denote the case with $\Ra_2$. The orange circular markers are the results obtained with PINNs, while the blue square markers are for the results obtained via nudging (extracted from \cite{agasthya_reconstructing_2022}). Both figures share the same legend.}
    \label{fig:kl_scan}
\end{figure}

To put matters into quantitative terms, in Fig.~\ref{fig:kl_scan} we show $\Delta_T$ and $\Delta_v$ obtained for the different $\ell/\delta$ and the two  Rayleigh numbers. The figures also show the results obtained via Nudging presented in \cite{agasthya_reconstructing_2022}. As shown  on the top panel of Fig.~\ref{fig:kl_scan} for the case of the temperature field, PINNs and Nudging perform similarly. On the other hand, concerning the most interesting -and difficult- question of reconstructing the velocity field, in the bottom panel we show that PINNs outperform Nudging when the supplied temperature is dense enough in space $\delta/\ell \sim 1$, while it is comparable with Nudging for sparse data, $\delta/\ell < 8\times 10^{-2}$. 

\begin{figure}
    \includegraphics[width=0.45\textwidth]{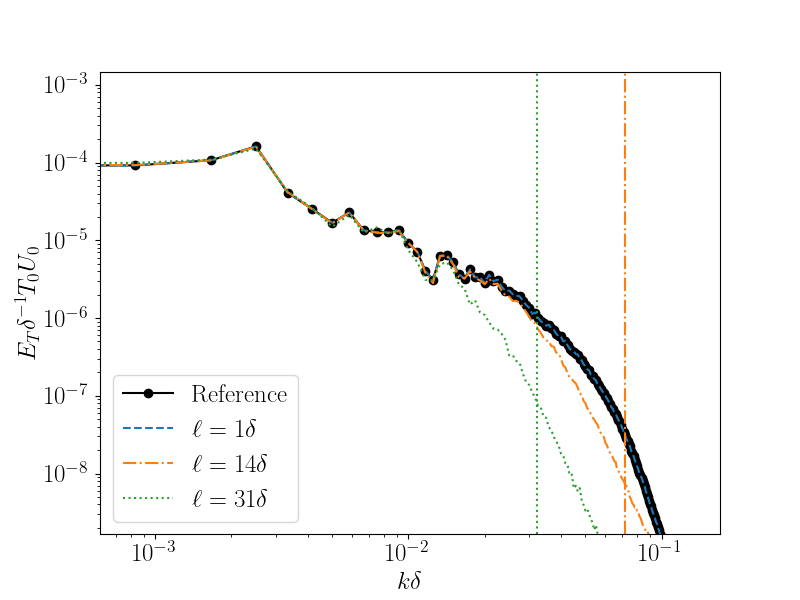}
    \includegraphics[width=0.45\textwidth]{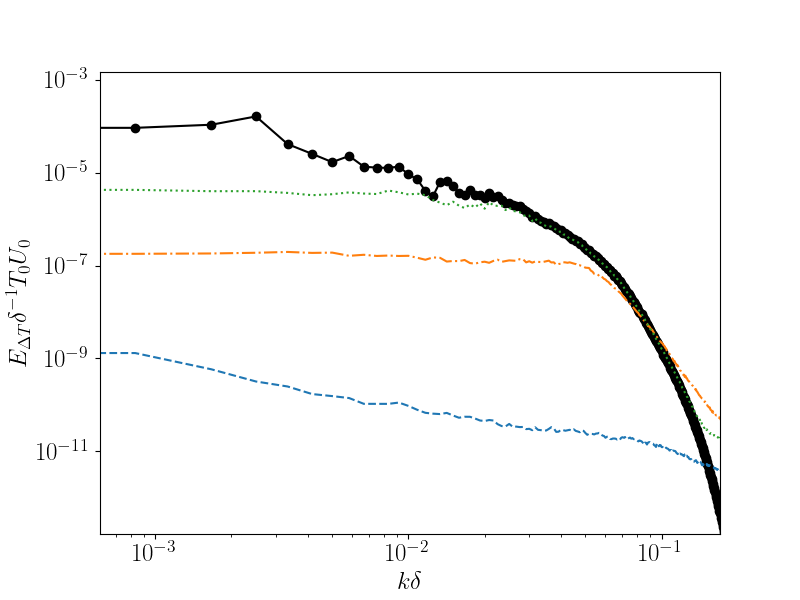}
    \caption{Top: temperature spectra for the reference flow at $\Ra_2$ and the spectra reconstructed fields obtained with $\ell/\delta=1$, 14, and 31. The vertical dash-dotted red line marks where $k\delta=14$, while the vertical dotted green line where $k\delta=31$. Bottom: temperature spectra for the reference flow at $\Ra_2$ and the spectra of the temperature error fields obtained with $\ell/\delta=1$, 14, and 31. Both figures share the same legend.}
    \label{fig:spectra_t}
\end{figure}

\begin{figure}
    \includegraphics[width=0.45\textwidth]{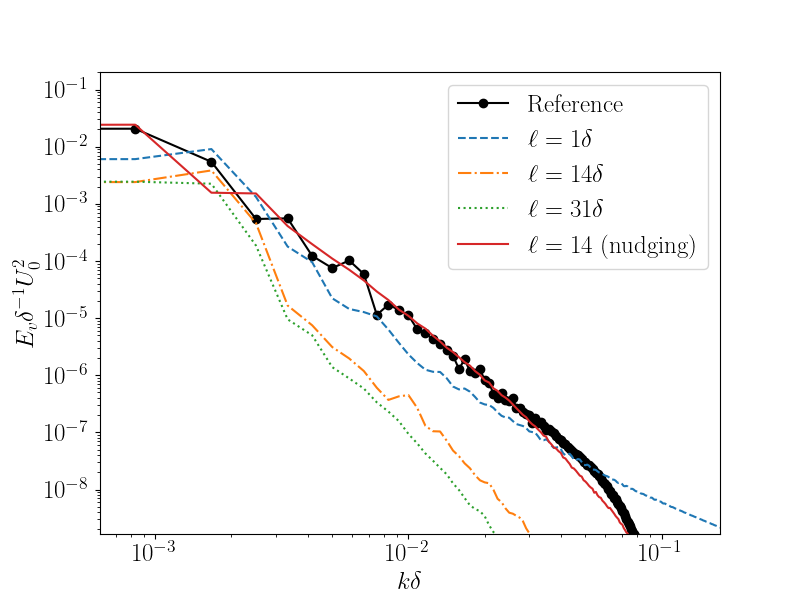}
    \includegraphics[width=0.45\textwidth]{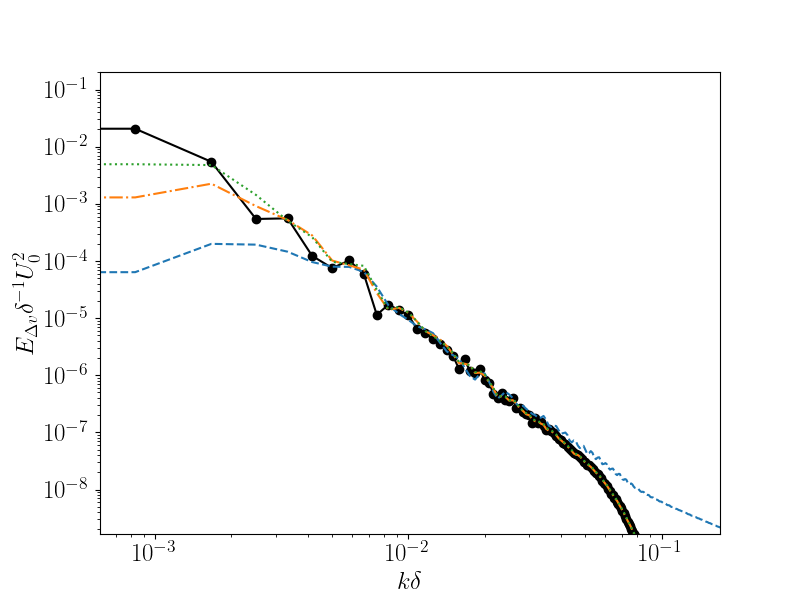}
    \caption{Top: vertical velocity spectra for the reference flow at $\Ra_2$ and the spectra reconstructed fields obtained with $\ell/\delta=1$, 14, and 31. The solid red line corresponds to the obtained via nudging \cite{agasthya_reconstructing_2022}. Bottom: vertical velocity spectra for the reference flow at $\Ra_2$ and the spectra of the vertical velocity error fields obtained with $\ell/\delta=1$, 14, and 31. Both figures share the same legend.}
    \label{fig:spectra_v}
\end{figure}

In Fig.~\ref{fig:spectra_t} (top) we show the temperature spectra at $y_0=300 \delta$ of the reference flow and of the reconstructions obtained with $\ell=1 \delta$, $14 \delta$, and $31 \delta$ for the high Rayleigh number case, while in the bottom panel we show the the spectra of the reference field superimposed with the spectra of the error $T_\Delta$. A Hanning window was applied to the reconstructed fields to cure any effects of non-periodicity. The vertical dash-dotted red line marks where $k\delta=14$ and the veritcal dotted green line where $k\delta=31$. As expected from the previous analysis, the overall errors are small and mostly concentrated in the small scales, although in the $\ell=14\delta$ and $31\delta$ cases, these are still bigger than the scale set by $\ell$. Figure~\ref{fig:spectra_v} shows the corresponding spectra of the vertical velocity, with the addition of the result obtained via nudging \cite{agasthya_reconstructing_2022} for $\ell/\delta=14$. Here, only in the case with $\ell/\delta=1$ the PINN produces a physically meaningful energy spectrum, while the spectra obtained via nudging is very similar to the reference one. In accordance with Figs.~\ref{fig:viz} and \ref{fig:prof_ra2}, the cases with higher $\ell/\delta$ can only properly reconstruct the position and shape of the largest structures of the flow, but fail to produce the small-scale structures observed at these Rayleigh numbers. The energy spectra of the reconstruction thus decay very rapidly.

\begin{figure}
    \includegraphics[width=0.45\textwidth]{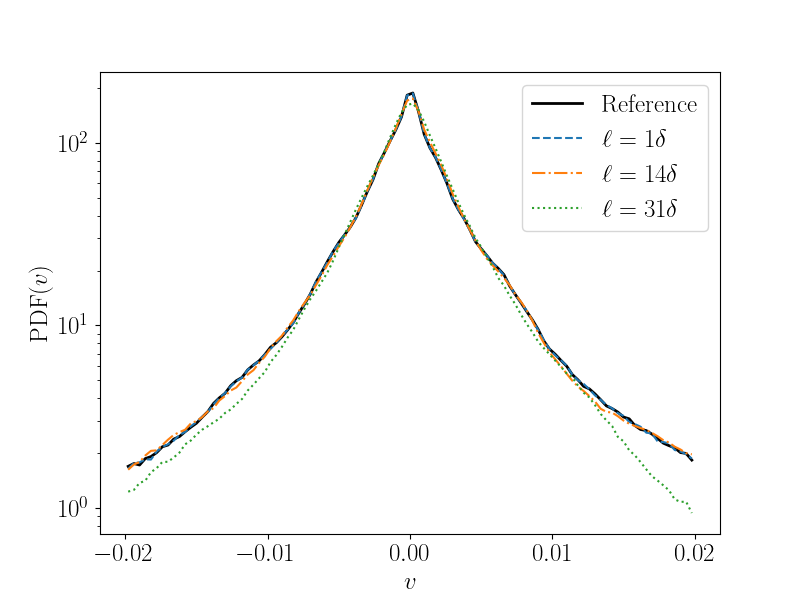}
    \includegraphics[width=0.45\textwidth]{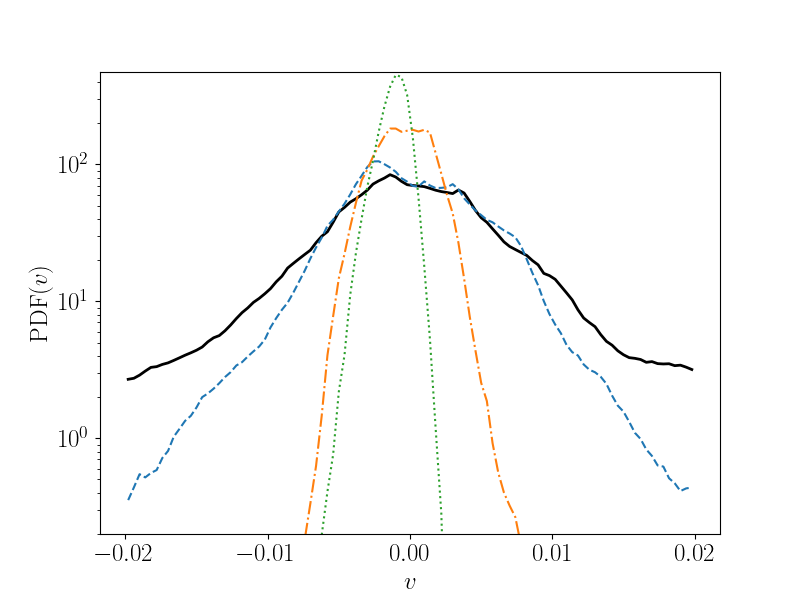}
    \caption{Temperature (top) and vertical velocity (bottom) probability density functions for the reference flow at $\Ra_2$ and for the reconstruction performed with $\ell/\delta=1$, 14, and 31. Both figures share the same legend.}
    \label{fig:pdfs}
\end{figure}

Finally, in Fig.~\ref{fig:pdfs}(a) and (b) we show the probability density functions (excluding the regions close to the walls) of temperature and vertical velocity, respectively for the case with $\Ra_2$. As expected, temperature statistics are well reproduced for all $\ell$ presented, while velocity statistics are only close to accurate when $\ell=1$, otherwise the probability density function start resembling a Gaussian.

\section{Conclusions}
\label{sec:conclusions}

Machine and deep learning techniques are becoming well-established tools in the Data Assimilation world. We show that it is possible to reconstruct Rayleigh-B\'enard  velocity fields having access only to time-resolved point-wise temperature measurements using PINNs. We investigate two different Rayleigh numbers at $7.2\times 10^7$ and $3.6 \times 10^8$. In order to assess the accuracy of PINNs we compare the results against a baseline given by Nudging \cite{agasthya_reconstructing_2022}. PINNs achieve an accuracy comparable to Nudging for the smallest Rayleigh number and a better accuracy for  highest Rayleigh case  when the temperature data are supplied at high spatial frequencies. On the other hand, when temperature is too sparse, PINNs fail to produce meaningful results, while Nudging still produces physically valid solutions. We interpret this as a lack to enforce the correct physical constraints when data are not dense enough. As seen in other works \cite{clark_di_leoni_reconstructing_2022,cuomo_scientific_2022}, PINNs can struggle with high-frequency components, a major problem in multi-scale turbulent data where information is key also at small-scales and high-frequencies.  It is  important to remark that while the results presented can be considered promising, we do not claim they are optimal under any criteria, better results may be obtained with a different choice of hyperparameters, a slight modifications to the architecture \cite{du_VarPINN2022}, by enforcing periodicity in the horizontal  direction via Fourier features \cite{tancik_fourier_2020}, or by splitting the domain into smaller subsections \cite{karniadakis_extended_2020}. It is out of the scope of this work to perform an exhaustive and meaningful hyperparameter scan. This work can be considered another exploratory attempt to systematic  assess  ML tools for reconstructing multi-scale turbulent fields, imposing physics constraints. 
Our intention here is to stress the importance to compare with other baselines (here Nudging is used), the need -in future works- to explore a wider range of Rayleigh (here limited to $[10^7:10^8]$) and to jump to full $3d+1$ space-time domains, the need to distinguish point-based reconstructions (here assessed in term of $L_2$ norm, and spectral errors)  from statistical
reconstruction (here assessed with spectral and probability density functions). As far as this study shows, when data are sparse, PINNs performance worsens. Contrary to Nudging, the obtained reconstructed fields not only have a high point-to-point error but also have incorrect energy spectra and statistics.

\section*{Declarations}

\noindent The datasets generated during and/or analysed during the current study are available from the corresponding author on reasonable request.

\noindent L.A. and P.C. conceived and carried out the numerical experiments, and analyzed
the results. All authors worked on developing the main idea, discussed the
results and contributed to the final manuscript.

\noindent This project has
received partial funding from the European Research Council
(ERC) under the European Union’s Horizon 2020 research
and innovation programme (Grant Agreement No. 882340)).

\bibliographystyle{ieeetr}

\bibliography{sn-bibliography}

\end{document}